\documentclass[sigconf,natbib=false]{acmart}

\AtBeginDocument{%
  \providecommand\BibTeX{{%
    \normalfont B\kern-0.5em{\scshape i\kern-0.25em b}\kern-0.8em\TeX}}}

    \setcopyright{none}
\renewcommand\footnotetextcopyrightpermission[1]{} 
\usepackage{graphicx}
\usepackage{amsmath}
\usepackage{booktabs}
\usepackage{epsfig}
\usepackage{enumitem}
\setlist[itemize]{leftmargin=3mm}
\usepackage{multirow}
\usepackage{makecell}
\usepackage{xr}
\usepackage{enumitem}
\usepackage{tikz}
\usepackage[framemethod=tikz]{mdframed}

\usepackage{hyperref}
\usepackage{dsfont}
\usepackage{subfigure}
\usepackage[ruled,vlined,linesnumbered]{algorithm2e}
\usepackage{algorithm2e}
\usepackage{algorithmicx}  



\def\eg{\emph{e.g.,}\xspace}

\def\ie{\emph{i.e.,}\xspace}
\def\etal{\emph{et al.}\xspace}

\begin{document}

\title{Towards Evaluating the Robustness of Automatic Speech Recognition Systems via Audio Style Transfer}

\author{Weifei Jin}
\email{weifeijin@bupt.edu.cn}
\affiliation{
  \institution{Beijing University of Posts and Telecommunications}
  \country{China}}

\author{Yuxin Cao}
\email{caoyx21@mails.tsinghua.edu.cn}
\affiliation{
  \institution{Tsinghua University}
  \country{China}
}

\author{Junjie Su}
\email{sujunjie@bupt.edu.cn}
\affiliation{
  \institution{Beijing University of Posts and Telecommunications}
  \country{China}}

\author{Qi Shen}
\email{shenqi629@bupt.edu.cn}
\affiliation{
  \institution{Beijing University of Posts and Telecommunications}
  \country{China}
}

\author{Kai Ye}
\email{yek21@mails.tsinghua.edu.cn}
\affiliation{
 \institution{Tsinghua University}
 \country{China}}

\author{Derui Wang}
\email{derui.wang.edu@gmail.com}
\affiliation{
  \institution{CSIRO's Data61}
  \country{Australia}}

\author{Jie Hao}
\authornote{Corresponding author.}
\email{haojie@bupt.edu.cn}
\affiliation{
  \institution{Beijing University of Posts and Telecommunications}
  \country{China}
  }

\author{Ziyao Liu}
\email{liuziyao@ntu.edu.sg}
\affiliation{
  \institution{Nanyang Technological University}
  \country{Singapore}}

\begin{abstract}
In light of the widespread application of \textbf{A}utomatic \textbf{S}peech \textbf{R}ecogni-tion (ASR) systems, their security concerns have received much more attention than ever before, primarily due to the susceptibility of Deep Neural Networks. Previous studies have illustrated that surreptitiously crafting adversarial perturbations enables the manipulation of speech recognition systems, resulting in the production of malicious commands. These attack methods mostly require adding noise perturbations under $\ell_p$ norm constraints, inevitably leaving behind artifacts of manual modifications. Recent research has alleviated this limitation by manipulating style vectors to synthesize adversarial examples based on Text-to-Speech (TTS) synthesis audio. However, style modifications based on optimization objectives significantly reduce the controllability and editability of audio styles. In this paper, we propose an attack on ASR systems based on user-customized style transfer. We first test the effect of \textbf{S}tyle \textbf{T}ransfer \textbf{A}ttack (STA) which combines style transfer and adversarial attack in sequential order. And then, as an improvement, we propose an iterative \textbf{S}tyle \textbf{C}ode \textbf{A}ttack (SCA) to maintain audio quality. Experimental results show that our method can meet the need for user-customized styles and achieve a success rate of 82\% in attacks, while keeping sound naturalness due to our user study.
\end{abstract}

\maketitle

\section{Introduction}

\begin{figure}[t]
  \centering
  \includegraphics[width=0.46\textwidth]{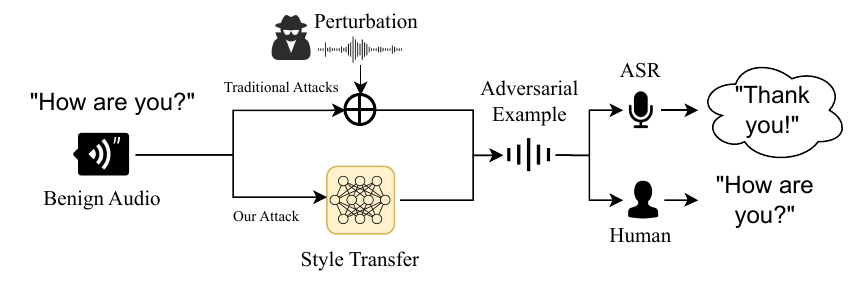} 
  \caption{Comparison between our attack and traditional attacks.}
  \label{fig:attack}
\end{figure}

\begin{figure*}[t]
  \centering
  \includegraphics[width=0.98\textwidth]{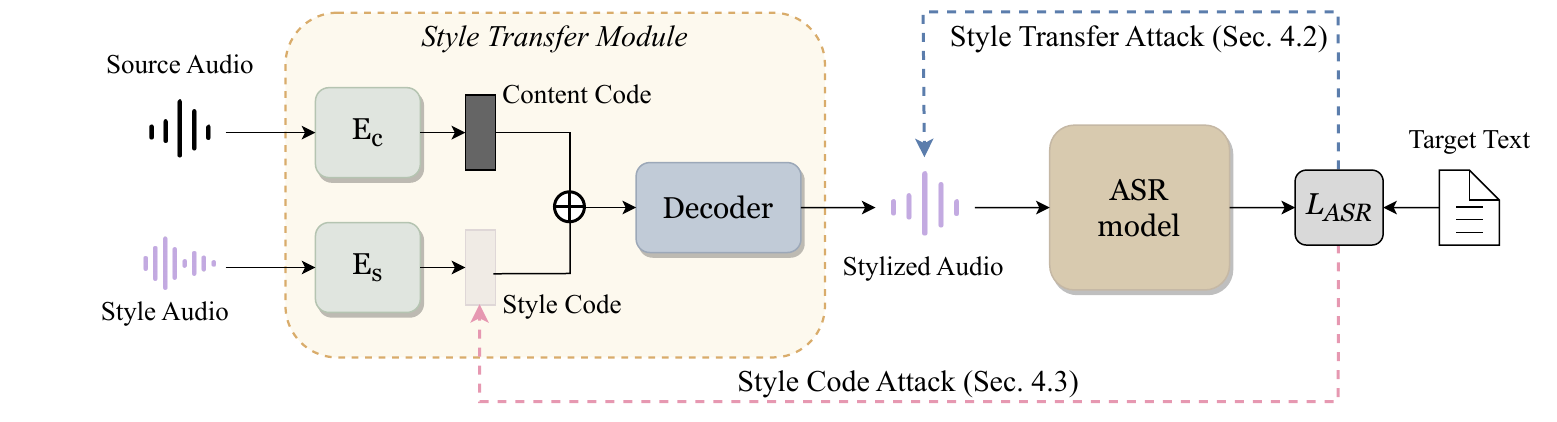} 
  \caption{An overview of STA and SCA.}
  \label{fig:overview}
\end{figure*}

It has become widely acknowledged that Deep Neural Networks (DNNs) are vulnerable to adversarial attacks~\cite{dai2018adversarial}, indicating that slight perturbations can cause models to make errors or degrade in performance without altering semantics. In the speech domain, Automatic Speech Recognition (ASR) systems are extensively used~\cite{wang2019overview,mehrish2023review}, embedded in various smart devices such as voice assistants~\cite{poushneh2021humanizing,yan2022survey}, and widely employed in providing subtitles for short videos which have enjoyed high popularity in the last decade~\cite{subtitle}. Given these realities, the potential harm posed by adversarial attacks on ASR models is significant. For instance, attackers could use  well-crafted adversarial examples to mislead voice assistants, causing them to execute malicious commands. Therefore, exploring potential attacks that ASR models may face is urgently needed and essential. 

In recent years, a lot of research has been devoted to investigate adversarial attacks on ASR models~\cite{li2020advpulse,kwon2019selective,zheng2021black}. Most of these works involve adding perturbations to benign audio to generate adversarial examples that deceive ASR models. The underlying premise is to limit the magnitude of perturbations to preserve semantics, with small perturbations being imperceptible to humans. However, these perturbation-based methods inevitably leave traces of manual modifications, which can be exploited to better defend against such attacks~\cite{hussain2021waveguard,yang2018characterizing,du2020unified}. 

To maintain the imperceptibility and audio quality of adversarial examples, recent works have loosen the traditional constraints on perturbations in adversarial attacks, thus evading defenses detecting artifacts from perturbations limited by $\ell_p$ norm constraints. One study employs Text-to-Speech (TTS) synthesis to generate adversarial examples by iteratively inputting style vectors into the TTS model to achieve attack objectives, resulting in significant changes in the audio waveform of the generated examples~\cite{qu2022synthesising}. Another study, SMACK~\cite{yu2023smack}, manipulates prosody vectors at a fine granularity and then similarly uses a TTS model to introduce semantically meaningful adversarial audio examples. However, directly iterating over style vectors based on optimization objectives results in generated audio styles that are not under user control. 

To address the limitation of controllability of audio attacks and explore the potential attacks in the aforementioned conditions, we propose a user-customized style transfer attack, enabling users to select desired styles and generate corresponding adversarial examples, which are not only harder to defend against and perceive but also support user-customized styles.
The difference between our attack and traditional attacks is illustrated in Figure~\ref{fig:attack}. 
Concretely, our attack considers the following application scenario. To avoid re-recording videos due to minor flaws and improve the efficiency of video creation, content creators on short video platforms, such as podcasts, TikTok, and YouTube, have diverse preferences for the style of their speech and typically do not want to change their voice. Then, the attacker can masquerade as a provider of speech style transfer services, which can not only customize speech styles based on user preferences that conveniently align with their needs, but also transform normal user-input speech into adversarial examples through the provided speech style transfer service. These examples may then produce undesirable outcomes in downstream ASR tasks when the videos are released online.

Specifically, we propose two attack schemes, \textbf{S}tyle \textbf{T}ransfer \textbf{A}ttack (STA) and \textbf{S}tyle \textbf{C}ode \textbf{A}ttack (SCA), as illustrated in Figure~\ref{fig:overview}. STA involves concatenating style transfer with adversarial attacks. The input audio undergoes style transfer before perturbations are added. This approach leverages the characteristics of style transfer to mitigate certain constraints of traditional adversarial attacks, making it more difficult to be defended. However, this method still inevitably introduces noise artifacts. To address this limitation, we apply an iterative optimization process to the style transfer model internally, proposing SCA. The basic principle is as follows: a speech segment is transformed into different style codes after passing through various style encoders. The decoder then generates the final audio based on these style codes, combined with content encoding. Each style code corresponds to the style of the generated audio. By perturbing these style codes, we can ensure that the synthesized audio still exhibits the style corresponding to the initial style audio, while potentially causing transcription errors by the ASR model, thus achieving the attack objective.

Our contributions are summarized as follows:
\begin{itemize}
    \item We propose a new adversarial audio attack method based on style transfer, which not only breaks the traditional adversarial perturbation constraint but also supports customized styles.
    \item We utilize style encoders to extract style information and perturb style codes, ensuring that the style remains unchanged while improving the naturalness and quality of the adversarial examples.
    \item The experimental results demonstrate the effectiveness and feasibility of our attack. Our user study also shows that our adversarial audio examples are indeed relatively more natural and harder to perceive by humans.
\end{itemize}

\section{Related Work}
In this section, we first briefly introduce audio style transfer and its development in recent years, followed by an overview of audio adversarial attacks against intelligent speech systems.

\subsection{Audio Style Transfer}
Audio style transfer aims to transform the style of an audio clip to match that of a target audio clip while preserving the semantic content of the original audio. Similar to image style transfer such as ~\cite{Gatys_2016_CVPR}, Grinstein \etal~\cite{grinstein2018audio} employed a style loss function to iteratively adjust the target audio, gradually aligning its style with that of the input style audio, thereby achieving style transfer in the end. AutoVC~\cite{qian2019autovc} is a voice conversion system utilizing an autoencoder architecture capable of achieving zero-shot audio style transfer. Later, SpeechSplit~\cite{qian2020unsupervised} effectively decomposes a speech into four components: content, pitch, rhythm, and timbre. SpeechSplit2.0~\cite{chan2022speechsplit2} alleviates the laborious bottleneck tuning through advanced speech signal processing techniques. Recently, numerous text-to-speech (TTS) models have achieved high-quality style transfer~\cite{lee21h_interspeech,chen2022fine,huang2022generspeech,li2022styletts}.

\subsection{Audio Adversarial Attack}
Audio adversarial examples~\cite{abdullah2021sok,zhang2022adversarial} are generated to deceive intelligent speech systems such as ASR models~\cite{wang2019overview}. Carlini and Wagner implemented targeted attacks against ASR models for arbitrary-length text~\cite{carlini2018audio}. Subsequently, Taori \etal~\cite{taori2019targeted} extended this to a black-box setting using genetic algorithms. Afterwards, many studies focused on aspects such as the robustness\cite{zhang2021generating,ijcai2019p741,esmaeilpour2022towards}, generation speed\cite{li2020advpulse,chiquier2022realtime,xie2021enabling}, and perceptibility\cite{guo2022specpatch,schonherr2018adversarial,qin2019imperceptible,abdullah2019practical,wu2023kenku} of adversarial examples. 
Qu \etal\cite{qu2022synthesising} were the first to use TTS models to generate audio adversarial examples without requiring source audio input. A recent work, SMACK~\cite{yu2023smack}, achieved semantically meaningful audio adversarial attacks by modifying prosody vectors. Different from the aforementioned approaches, our method not only improved the naturalness of the audio while reducing the perceptibility of the generated examples, but also enables the modification of audio into user-customized styles.

\section{Threat Model}
\noindent\textbf{Attacker's Objective.} Given an ASR model, the attacker's objective is to conduct targeted attacks in the digital domain. This is achieved by generating an audio adversarial example, causing its transcription by the ASR model to be inconsistent with the corresponding ground truth text. To minimize human detection, the attacker aims to enhance the quality and naturalness of the generated adversarial example while ensuring its content remains unchanged.

\noindent\textbf{Attacker's Background Knowledge.} Consistent with previous work~\cite{qu2022synthesising,zhang2021generating}, we operate under a white-box setting, wherein the attacker has access to the model architecture and parameter information. This allows the attacker to directly compute gradients through back-propagation and update their optimization objective.

\noindent\textbf{Attack Scenario.} We assume that the attacker masquerades as a speech style transfer service provider, which can help users customize the style of speech modifications, such as adjusting the speech speed or modifying certain parts of a singing recording's pitch. This is applicable to many everyday scenarios, such as students who submit audio assignments, teachers who record lecture videos, or social media influencers who share food and fashion content. In these cases, users typically do not want their voices to sound like someone else's, indicating that the timbre should not change. The speech style transfer service disguised by the attacker can fulfill this requirement. However, the audio generated through the service provided by the attacker can also carry imperceptible adversarial noise, which can deceive ASR models into making transcription errors. These examples, when fed into ASR models (\eg automatic subtitles for videos), may leak false information. Once these examples are disseminated through the network, they can lead to negative consequences.

\section{Methodology}
In this section, we first provide the definition of the optimization problem for generating adversarial examples against ASR. Then, we elaborate on the two proposed attack schemes.

\subsection{Problem Definition}
Given a well-trained ASR model $f$ and an input audio clip $x$, the attacker's objective is to optimize the following loss function.
\begin{equation}
\begin{aligned}
\mathop {\min }\limits_{x'} \quad & {L_{ASR}}\left( {f\left( {x' } \right),t} \right), & \\
\mbox{s.t.}\quad
& f\left( {x' } \right) = {t}, g\left(x'\right)=g\left(x\right), &
\end{aligned}
\end{equation}
where $x'$ stands for the adversarial audio that the attacker hopes to find, $t$ denotes the target text that is different from $t$, and that the attacker expects the ASR model to transcribe,  $g$ represents the transcription result perceived by humans. 

For $L_{ASR}$, following prior work~\cite{carlini2018audio,qu2022synthesising,zhang2021generating}, we introduce the Connectionist Temporal Classification (CTC) loss~\cite{graves2006connectionist}, which is commonly used for sequence labeling tasks, particularly suitable for scenarios without aligned labels, such as frame-level annotations in speech recognition. It aggregates across various alignments of the input sequence to find the most probable output sequence, thereby computing the loss. Let $F$ denote the probability distribution output by the ASR model, and let $D_g$ represent a decoder, such as a greedy decoder, which is used to decode the probability distribution $F\left(x\right)$ output by the ASR model into a transcription. Thus, $f\left(x\right) = D_g\left(F\left(x\right)\right)$. Then the CTC loss is calculated by: 
\begin{equation}
L_{CTC}\left(x,t\right) = -\log \sum_{\pi \in \Pi\left(t\right)}\prod_{i} F\left(x\right)_{\pi_i}^i,
\end{equation}
where $\Pi\left(t\right)$ denotes the alignment set derived from the targeted transcription $t$, $\pi$ is an alignment of $t$, satisfying that $\pi$ can be reduced to $t$, and the length of $\pi$ equals that of $t$, $F\left(x\right)_{\pi_i}^i$ denotes the probability of $\pi_i$ in the $i$-th frame.
\noindent
\begin{algorithm}[t]
\caption{STA}\label{alg:STA}
\KwIn{ASR model $f$, source audio ${x}$, style audio $x_s$, target text $t$, style transfer model $G$, Levenshtein Distance $D_L$, maximum iteration number $N$, intensity width threshold of the perturbation $b$, learning rate $\alpha$, weighting factor $\lambda$, loss function $L_{ASR}$.}
\KwOut{Adversarial audio $x_{STA}$.}
$z \gets G\left(x, x_s\right)$\;
$\epsilon \gets b \cdot |z|$\;
Initialize $w$ with random value from $N\left(0,1\right)$\;
\For {$i \gets 1$ to $N$}{
$\delta \gets \epsilon\cdot \tanh\left(w\right)$\;
$x' \gets z + \delta$\;
\If{$D_L\left(f\left(x'\right),t\right) == 0$}{
    $x_{STA}=x'$\;
    \textbf{break}
}
$l \gets L_{ASR}\left(f\left(x'\right),t\right) + \lambda||\delta||_2^2$\;
$w \gets w - \alpha \cdot {\nabla _w}l$\;
}
$x_{STA} \gets z + \epsilon\cdot \tanh\left(\delta\right).$
\end{algorithm}

\subsection{Style Transfer Attack}\label{sec:STA}

Influenced by the concepts of StyleFool~\cite{cao2023stylefool}, we initially integrate a style transfer step into the conventional audio adversarial attack pipeline. Given a style transfer model $G$, for an input audio $x$ and a style reference audio $x_s$, we first obtain a stylized audio $z = G\left(x, x_s\right)$, then apply a naive adversarial attack method to add perturbations to $z$. Formally, the attacker's objective can be expressed as: 
\begin{equation}
\begin{aligned}
\mathop {\min }\limits_{\delta} \quad & {L_{ASR}}\left( {f\left( {{z + \delta }} \right),t} \right), & \\
\mbox{s.t.}\quad
& f\left( {z + \delta } \right) = t, g\left( {z + \delta } \right) = g\left( z \right) = g\left(x\right), &
\end{aligned}
\end{equation}
where $\delta$ is the perturbation to be optimized which will be added to the stylized audio $z$.

Inspired by~\cite{zhang2021generating,tong2023query}, in order to better align the generated adversarial example with human auditory perception, we leverage Temporal Dependencies (TD), maintaining the proportional relationship between the perturbation $\delta$ and the stylized audio $z$. Specifically, the objective loss function for generating adversarial examples is as follows.
\begin{equation}\label{eq:frac_dlt}
\begin{aligned}
\mathop {\min }\limits_{\delta} \quad & {L_{ASR}}\left( {f\left( {{z + \delta }} \right),t} \right)+ \lambda \left\| \delta  \right\|_2^2, & \\
\mbox{s.t.}\quad
& \left| {\frac{\delta }{z}} \right| < b, &
\end{aligned}
\end{equation}
where $b$ denotes the intensity width threshold of the perturbation, which is within the range of $\left[ {0,1} \right]$.
$\lambda$ is the weighting factor between the adversarial objective and the similarity to the stylized audio, the regularization term is used to reduce noise. Considering that applying clip operation to local points may potentially impact the overall quality and in order to efficiently handle the constraints on $\delta$, we adopt the \textit{change of variable} method previously implemented in~\cite{carlini2017towards} by using the hyperbolic tangent function $\tanh \left( \cdot \right)$, thereby transforming Equation~\ref{eq:frac_dlt} into:
\begin{equation}
\begin{aligned}
\mathop {\min }\limits_{w} \quad & {L_{ASR}}\left( {f\left( {{z + \delta }} \right),t} \right)+ \lambda \left\| \delta  \right\|_2^2, & \\
\mbox{s.t.}\quad
& \delta  = b\left| z \right|\cdot\tanh \left( w \right), &
\end{aligned}
\end{equation}
where $w$ shares the same shape as $\delta$. With the transformation above, the variable to be optimized can take any value in the field of real numbers, and it can still ensure that the adversarial audio does not exceed the perturbation range, thus eliminating the need for clip operations. The algorithm details are shown in Algorithm \ref{alg:STA}.

\noindent
\begin{algorithm}[t]
\caption{SCA}\label{alg:SCA}
\KwIn{ASR model $f$, source audio ${x}$, style audio $x_s$, target text $t$, content encoder $E_c$, style encoder $E_s$, decoder $D$, Levenshtein Distance $D_L$, maximum iteration number $N$, learning rate $\alpha$, loss function $L_{ASR}$.}
\KwOut{Adversarial audio $x_{SCA}$.}
$c \gets E_c\left(x\right)$\;
$s \gets E_s\left(x_s\right)$\;
\For {$i \gets 1$ to $N$}{
$x'=D\left(c,s\right)$\;
\If{$D_L\left(f\left(x'\right),t\right) == 0$}{
    $x_{SCA}=x'$ \;
    \textbf{break}
}
$l \gets L_{ASR}\left(f\left(x'\right),t\right)$\;
$s \gets s-\alpha \cdot {\nabla _s}{l}$ \;
}
$x_{SCA} \gets D\left(c,s\right).$
\end{algorithm}

\subsection{Style Code Attack}\label{sec:SCA}
Through the aforementioned steps, the generated adversarial example $z'=z+\delta$ not only achieves the attack objective but also break the constraint on the variation between $z'$ and $z$ through style transfer operations. 
This characteristic significantly distinguishes this method from traditional audio adversarial attacks.
However, this approach does not effectively leverage the specific attributes and characteristics unique to speech and inevitably leaves behind artifacts of manual manipulation. To address this issue, we introduce the SCA method.

In this subsection, we leverage speech attributes such as content, pitch, rhythm, and timbre. While keeping the content fixed, we optimize the other attributes, which we refer to as speech style attributes, to generate adversarial examples.

\noindent\textbf{Style Modeling.}
Here, we utilize the advanced technique SpeechSplit2.0~\cite{chan2022speechsplit2}, which employs an autoencoder architecture, for style modeling. 
According to our threat model, users typically want to preserve their original timbre during style transfer, we do not consider timbre transfer and only focus on pitch and rhythm transfer.
We refer to the encoder used to encode speech style attributes, including pitch and rhythm, as the style encoder, denoted by $E_s$, the content encoder as $E_c$, and the decoder as $D$. The decoder takes content, pitch, rhythm, and timbre as inputs and generates the resulting speech spectrogram.

\begin{figure}[t]
  \centering
  \includegraphics[width=0.35\textwidth]{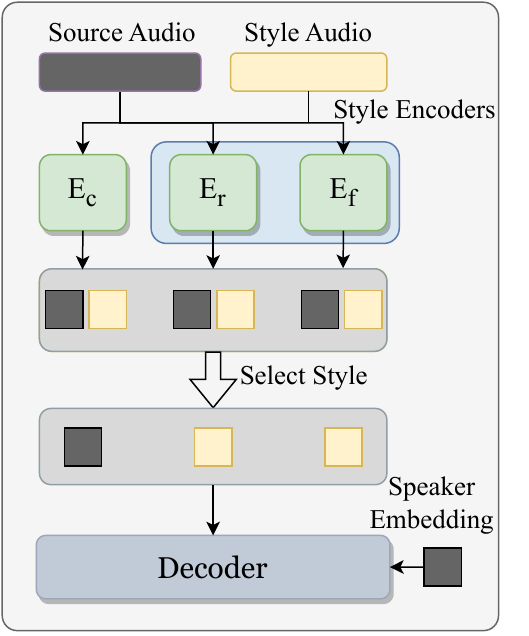} 
  \caption{Illustration of style transfer operations. There are three style selection options: pitch only, rhythm only, and both pitch and rhythm. For our attacks, we replace both pitch and rhythm code simultaneously.}
  \label{fig:autoencoder}
\end{figure}

\noindent\textbf{Style Code Optimization.} As is shown in Algorithm \ref{alg:SCA}, after passing through $E_c$ and $E_s$, both the source audio and the style audio are transformed into content code and style code. We first replace the style code of the source audio with that of the style audio to achieve style transfer. This process is illustrated in Figure~\ref{fig:autoencoder}. Then, we take the style code of the style audio as the variable to be optimized, and perform optimization using gradient descent. Specifically, our goal is to optimize the following expression.
\begin{equation}
{\mathop {\min }\limits_s {L_{ASR}}\left( {f\left( {D\left( {{E_c}\left( x \right),s} \right)} \right),t} \right)},
\end{equation}
where $s = {{E_s}\left( x_s \right)}$ denotes the style code of the style audio $x_s$.
The style code is then optimized by calculating the gradient ${\nabla _s}{L_{ASR}}$ until the transcribed text turns out to be the same as the target text.

Since the content code used in the synthesis results is derived from the source audio, our optimization operation does not alter the content of the resulting audio. 
Much to our surprise, optimizing the style code without constraints (due to its challenging definition) does not significantly impact the style of the resulting adversarial audio, allowing it to maintain the user-customized style. Please refer to Section~\ref{sec:discussion} for more detailed discussion. 

\section{Experimental Results}
In this section, we first analyze the attack performance of our methods, followed by our user study. Then we analyze the acoustic features of STA and SCA. Finally, we reveal the impact of different style codes on attack performance through ablation experiments.

\subsection{Experiment Settings}

\noindent\textbf{Dataset.} The VCTK corpus~\cite{yamagishi2019cstr} includes recordings from 110 English native speakers with diverse accents, comprising approximately 400 audio examples per speaker. In our experiment, we randomly select 100 audio examples, each featuring a target phrase. For each audio example, a style audio of a randomly chosen different speaker is selected. Each target phrase is generated by randomly replacing, inserting, or deleting each word in the phrase with different probabilities. As a result, we obtain pairs of ``source text-target text'' with different Levenshtein distances~\cite{yujian2007normalized}, allowing us to evaluate the feasibility of the attack. Levenshtein distance is a metric to measure the similarity between two text strings. It is calculated by determining the minimum number of operations required to transform one string into another, where inserting, deleting, or replacing a character counts as one operation.

\noindent\textbf{Model Settings.} Similar to ~\cite{guo2022specpatch,chiquier2022realtime,liu2020weighted}, we use DeepSpeech2~\cite{amodei2016deep} as our target ASR model. We build upon an open-source implementation~\footnote{https://github.com/SeanNaren/deepspeech.pytorch} and pre-train it on the VCTK dataset, resulting in a reduction of the Word Error Rate (WER) to 22.1 on the validation set. We also pre-train the style transfer model SpeechSplit2.0 on the VCTK dataset. For the STA method proposed in Section~\ref{sec:STA} and SCA method proposed in Section~\ref{sec:SCA}, we simultaneously transfer pitch and rhythm. 

\noindent\textbf{Metrics.} We adopt the following metrics to evaluate the performance of our attack methods.
\begin{itemize}
  \item \textbf{Success Rate (SR)} represents the ratio of successful adversarial examples $N_s$ to the total number of examples $N_t$ in the experiment, and is calculated by
  \begin{equation}
    SR = \frac{N_s}{N_t}.
  \end{equation}
  \item \textbf{Word Error Rate (WER)} is used to measure the difference between the target text and the decoded text, and is calculated by
  \begin{equation}
    WER = \frac{S+D+I}{N},
  \end{equation}
  where $S$, $D$ and $I$ respectively represent the number of substituted, deleted, and inserted words, $N$ is the total number of words.
\end{itemize}

\noindent\textbf{Parameter Settings.}
We set the maximum number of iterations to 3,000, meaning that any example that has not been successfully attacked by this number is considered a failure. Only examples in which every character is fully transcribed into the target text within the maximum number of iterations are considered successful attacks. For the gradient descent process, we use the Adam~\cite{DBLP:journals/corr/KingmaB14} optimizer with a learning rate set to 0.01. For STA, according to~\cite{zhang2021generating}, we set the intensity width threshold of the perturbation $b$ to 0.2.

\subsection{Attack Performance Analyses}
We first evaluate the performance of STA when simultaneously transferring pitch and rhythm and compare it with SCA when simultaneously iterating pitch code and rhythm code. The overall results are presented in Table~\ref{tab:results}. We respectively apply STA and SCA to simultaneously transfer pitch and rhythm, denoted as ``STA-pit+rhy'' and ``SCA-pit+rhy''. Both methods achieve a high success rate in the attacks. The results indicate that SCA achieves a success rate of 82\%, which is comparable to STA (85\%). This suggests that transitioning from perturbing the audio directly to iterating on the style code does not significantly reduce the attack success rate. Moreover, it also achieves better naturalness, which is discussed later in Section~\ref{sec:mos}. 

\begin{figure}[t]
  \centering
  \includegraphics[width=0.48\textwidth]{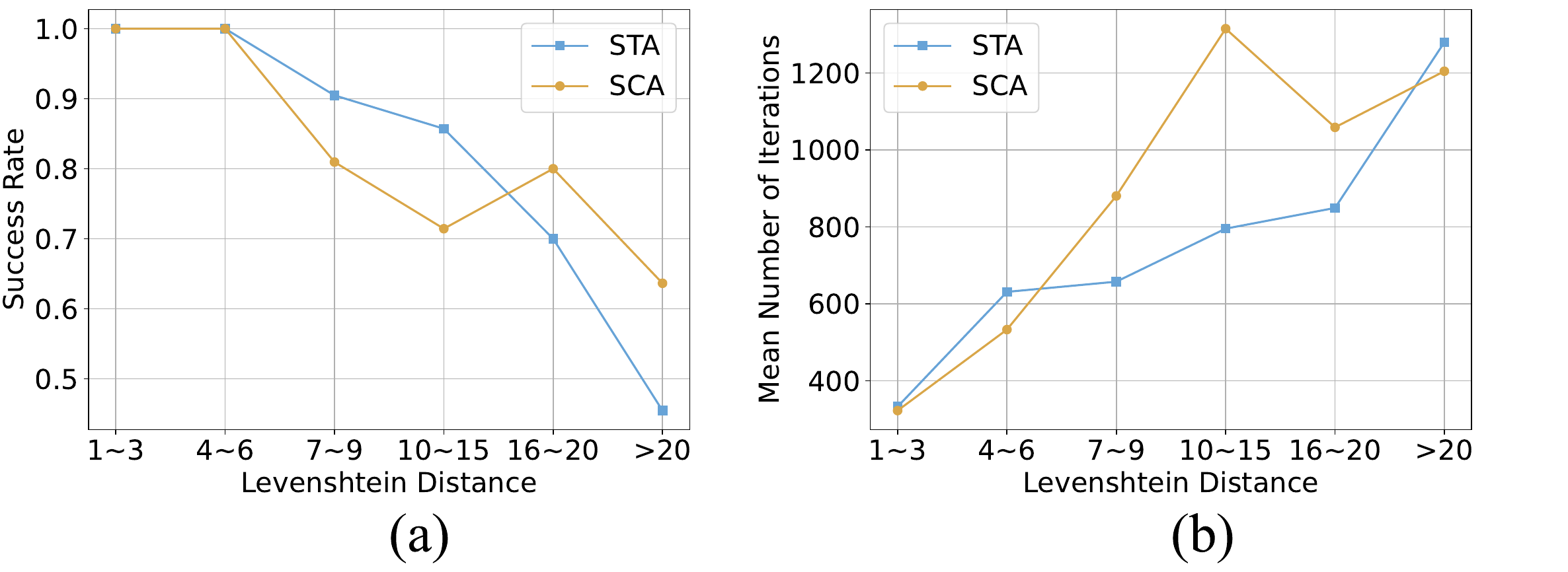}
  \caption{Performance of STA and SCA under different Levenshtein distances.
  }
  \label{fig:both_cmp}
\end{figure}

We conduct a further evalution to assess the success rate of both methods at different Levenshtein distances, which is shown in figure~\ref{fig:both_cmp}. We find that both SCA and STA exhibit good performance when the Levenshtein distance is small. However, when the Levenshtein distance increases beyond 6, SCA's attack performance deteriorates compared to STA. Specifically, SCA demonstrates a faster decline in attack success rate, accompanied by a more rapid increase in the required iteration count. For instance, within the Levenshtein distance range of 10 to 15, SCA achieves a success rate of only approximately 71.4\%, with an average iteration count reaching 1,316, while STA maintains a success rate of 85.7\% with an average iteration count of only 795. However, when the Levenshtein distance exceeds 20, SCA's performance surpasses that of STA. Overall, there is not much difference in the attack performance between the two methods.

\begin{figure}[t]
  \centering
  \includegraphics[width=0.42\textwidth]{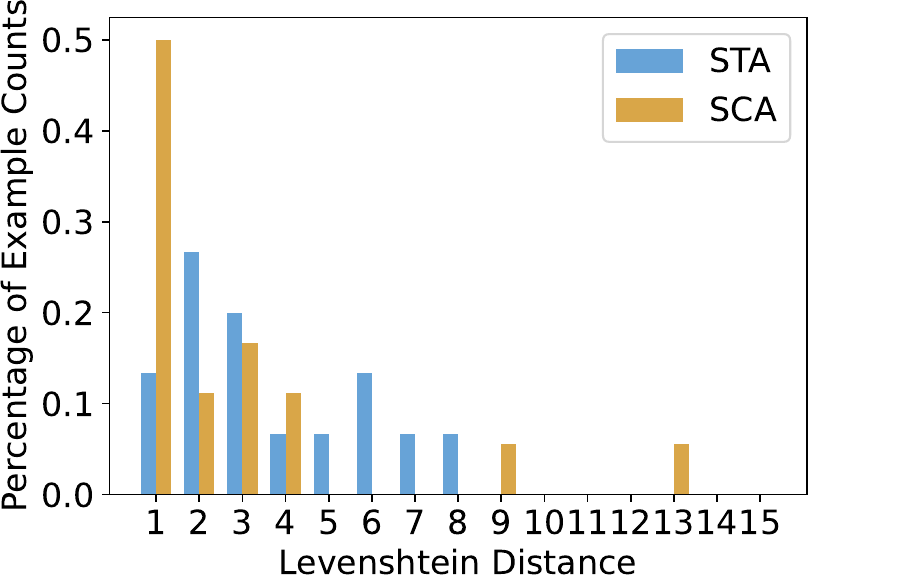} 
  \caption{The proportion of examples with the Levenshtein distance between the transcription result and the target text after the maximum number of iterations to the total number of failed attack examples.}
  \label{fig:remain_iter}
\end{figure}

For the failed attack examples, we compute the Levenshtein distance between their transcription results after the maximum number of iterations and the target text, as shown in Figure~\ref{fig:remain_iter}. The majority of these values are small, indicating proximity to success. Particularly for SCA, nearly 50\% of the failed examples have a Levenshtein distance of 1 from the target text, suggesting that this may primarily be influenced by the maximum iteration number. For STA, the Levenshtein distances between the transcription results of all failed examples and the target text are concentrated within the range of 1 to 8. Notably, neither STA nor SCA exhibits failed examples with large Levenshtein distances (exceeding 15). Overall, the difference in attack performance between the two methods is not significant.

\begin{table}[t]
  \centering
  \caption{Results for both STA and SCA under three style transfer schemes, as well as the ``Attack-only'' case.}
  \label{tab:results}
  \begin{tabular}{l|l|l}
    \toprule
    Attack methods & SR$\uparrow$ & WER$\downarrow$ \\
    \midrule
    Attack-only & 61\% & 17.8\% \\
    STA-pit & 68\% & 9.8\% \\
    STA-rhy & 71\% & 10.0\% \\
    \bfseries STA-pit+rhy & \bfseries 85\% & \bfseries 7.2\% \\
    SCA-pit & 8\% & 48.4\% \\
    SCA-rhy & 76\% & 7.6\% \\
    \bfseries SCA-pit+rhy & \bfseries 82\% & \bfseries 4.3\% \\
    \bottomrule
  \end{tabular}

\end{table}

\begin{figure}[t]
  \centering
  \includegraphics[width=0.45\textwidth]{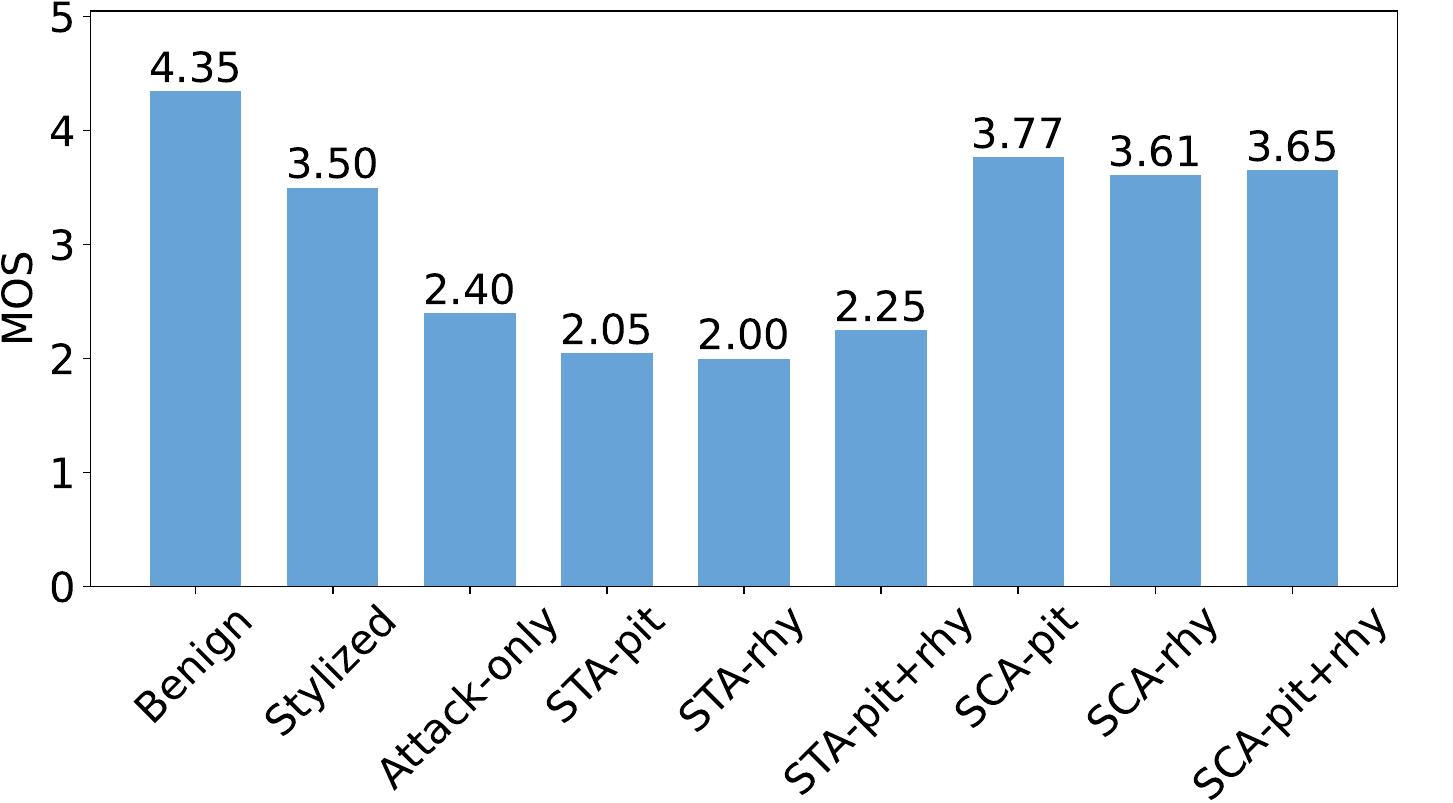} 
  \caption{Comparison of MOS for different audio sources.}
  \label{fig:MOS}
\end{figure}

\subsection{User Study}\label{sec:mos}
To evaluate the quality and naturalness of the audio adversarial examples generated by our methods, we designed a user study on Amazon Mechanical Turk~\cite{amazon_turk} and totally recruited 80 anonymous participants who are over 18 years old, and are able to complete the survey in English. We adhered to the best practice for ethical human subjects survey research, and all participants provided consent for their responses to be utilized for academic research purposes.

We use the Mean Opinion Score (MOS) to rate the speech quality, which is based on a Likert scale~\cite{likert1932technique} ranging from 1 to 5, where 1 indicates very poor quality, 2 indicates poor quality, 3 indicates neutral quality, 4 indicates good quality, and 5 indicates very good quality. We randomly selected four audio clips from each of the successful adversarial examples in the six main experiments shown in Table~\ref{tab:results}. Additionally, we selected four benign audio clips, four audio clips subjected only to style transfer, and four audio clips subjected only to the attack. 
Participants were asked to rate the quality of the audio according to their perception.
Furthermore, to further evaluate whether SCA affects the style of the resulting audio, we also selected four pairs of audio clips: one pair consisting of audio clips subjected only to style transfer and another pair consisting of audio clips subjected to SCA. Participants were asked to judge whether the pitch and rhythm of each pair of audio clips were the same. In total, our questionnaire consisted of 40 audio clips (or pairs), and the survey results are presented in Figure~\ref{fig:MOS}.

\begin{figure*}[t]
  \centering
  \subfigure[Source audio]{
    \includegraphics[width=0.31\textwidth]{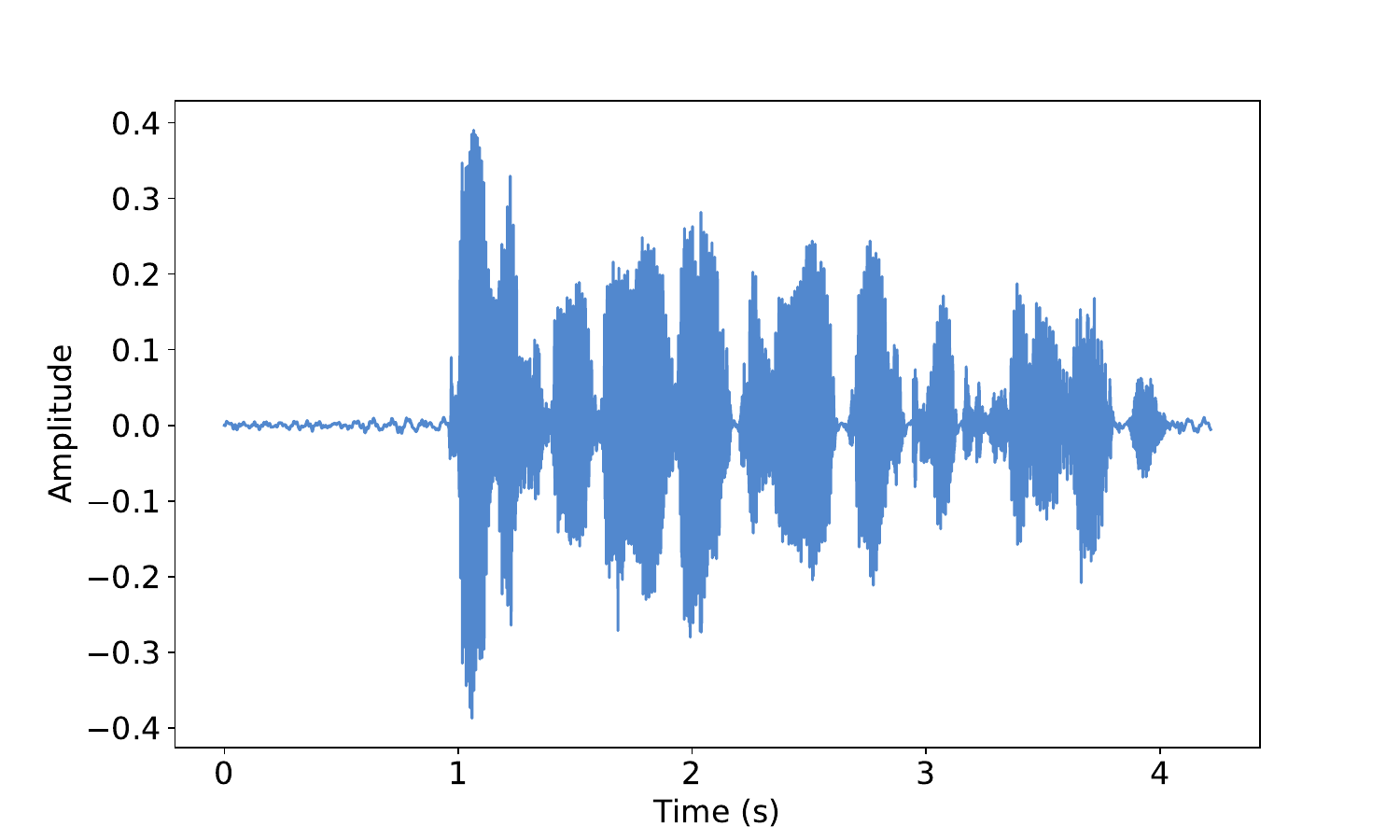}
    \label{fig:wav-a}
  }
  \subfigure[After style transfer]{
    \includegraphics[width=0.31\textwidth]{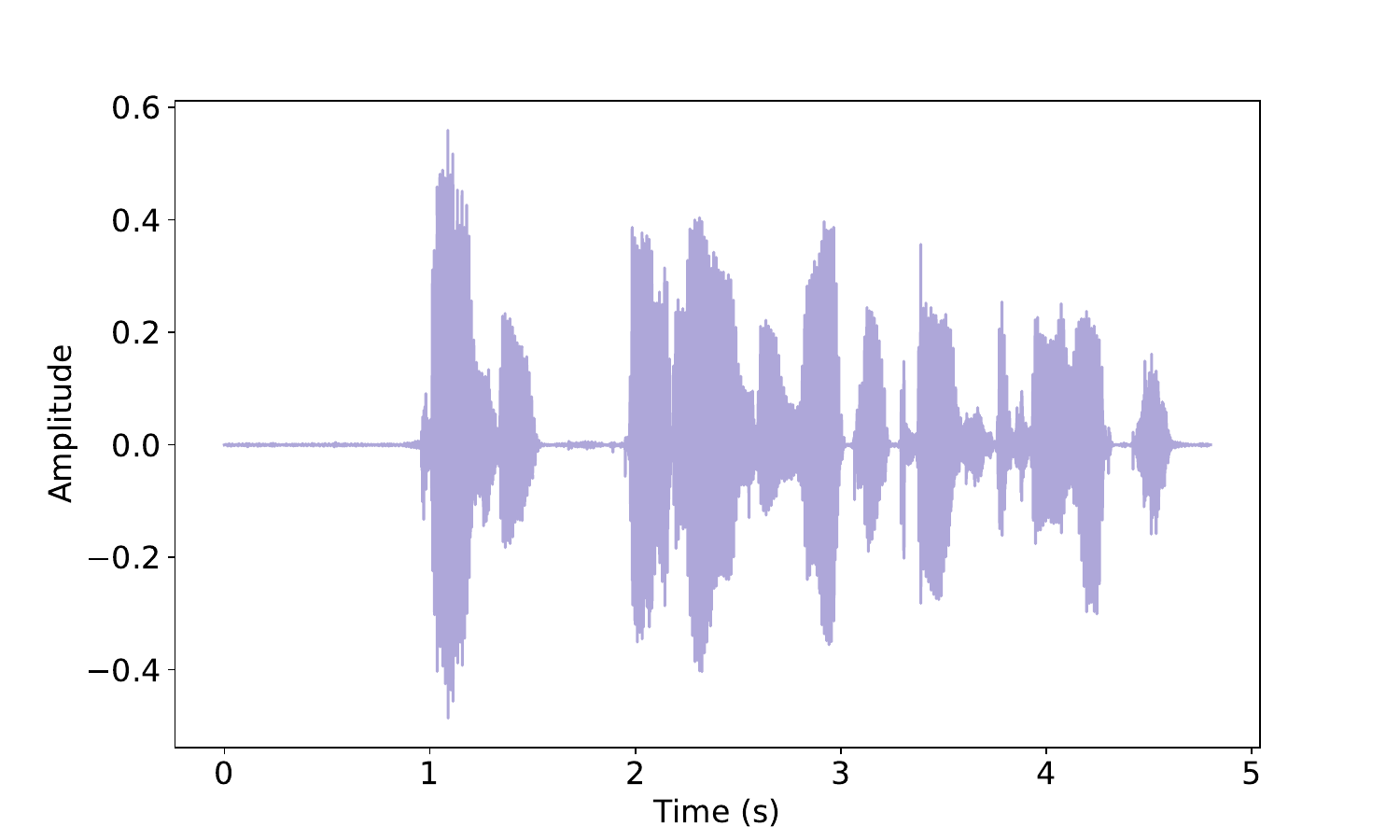}
    \label{fig:wav-b}
  }
  \subfigure[After STA]{
    \includegraphics[width=0.31\textwidth]{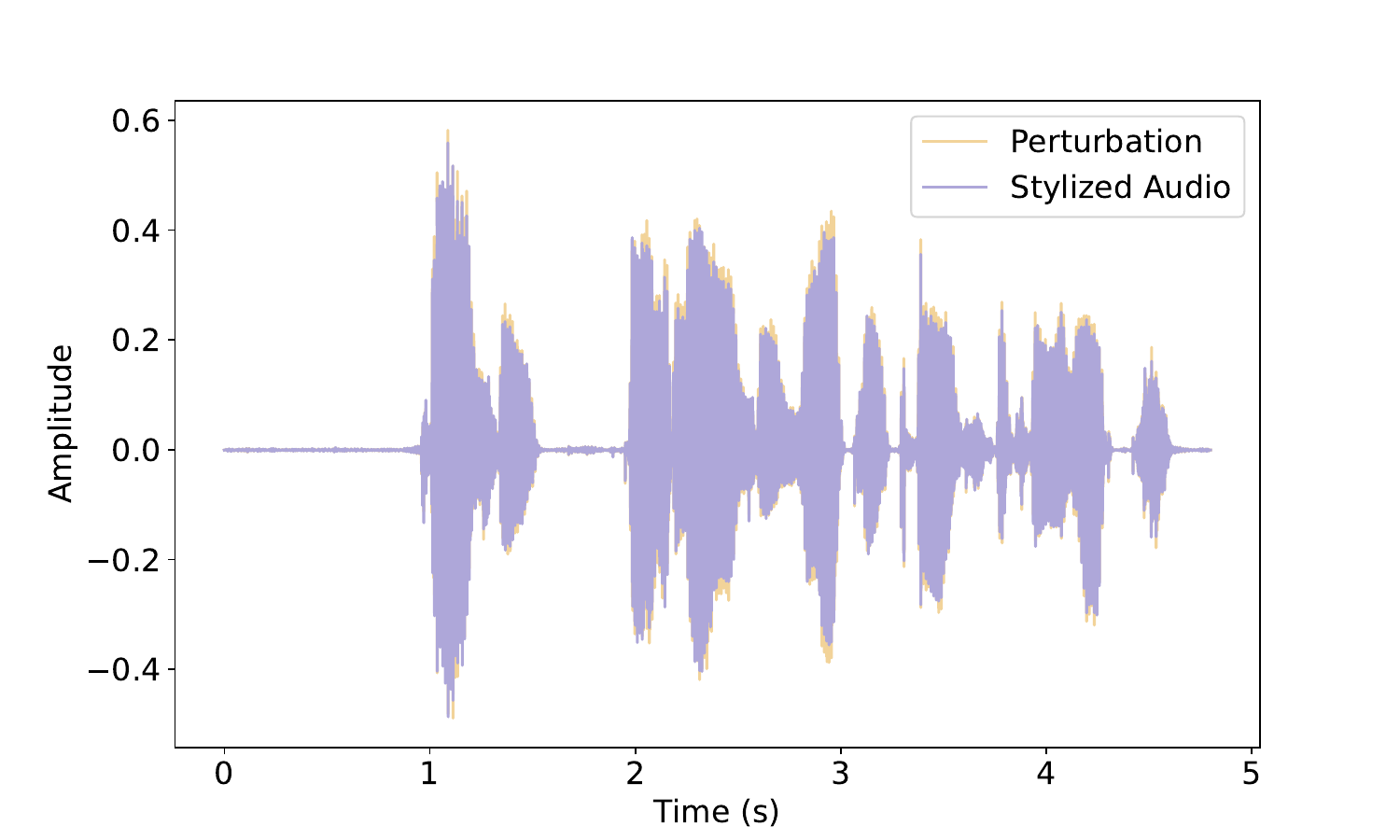}
    \label{fig:wav-c}
  }
  \caption{Waveforms of the source audio, the audio after style transfer, and the audio after STA.}
  \label{fig:wav}
\end{figure*}

\begin{figure}[t]
  \centering
    \subfigure[Source audio]{
    \includegraphics[width=0.22\textwidth]{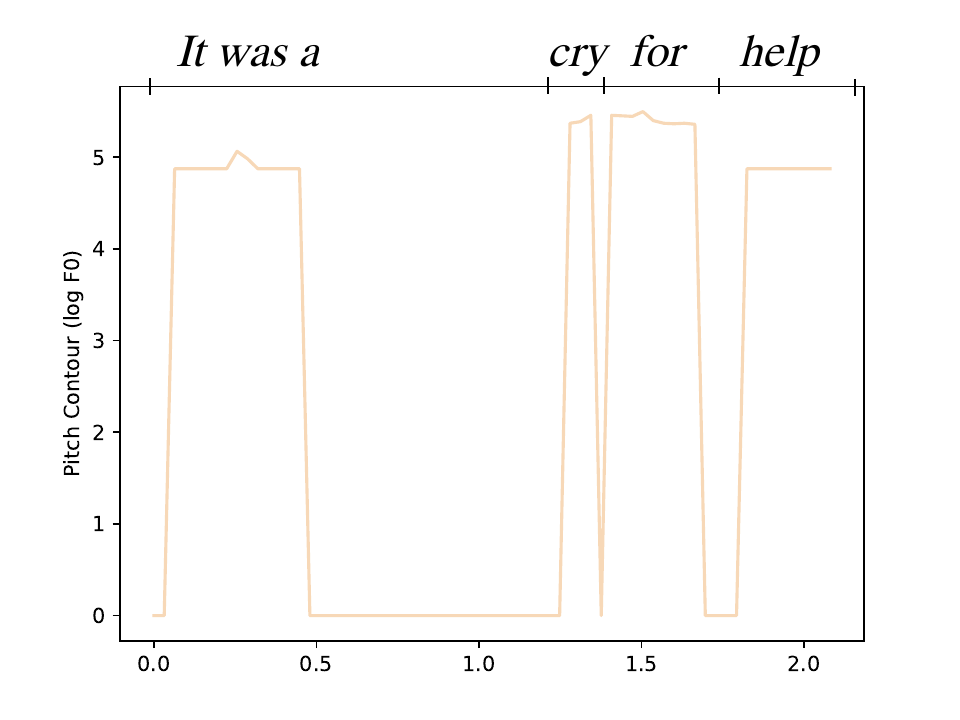}
    \label{fig:mel-a}
  }
  \subfigure[Style audio]{
    \includegraphics[width=0.22\textwidth]{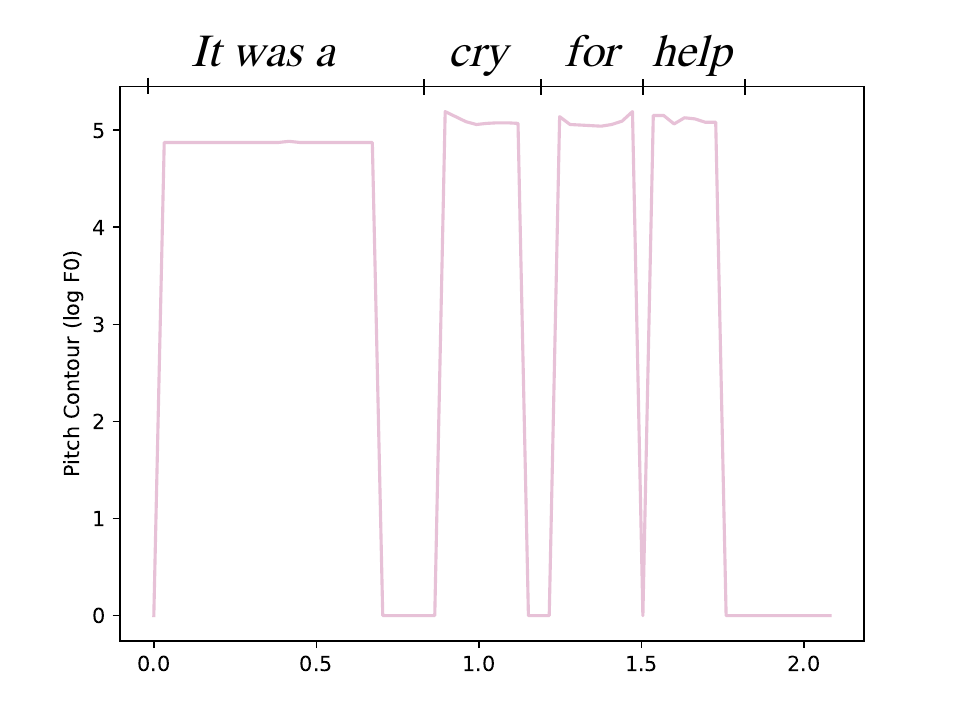}
    \label{fig:mel-b}
  }
  
  \subfigure[Stylized audio]{
    \includegraphics[width=0.22\textwidth]{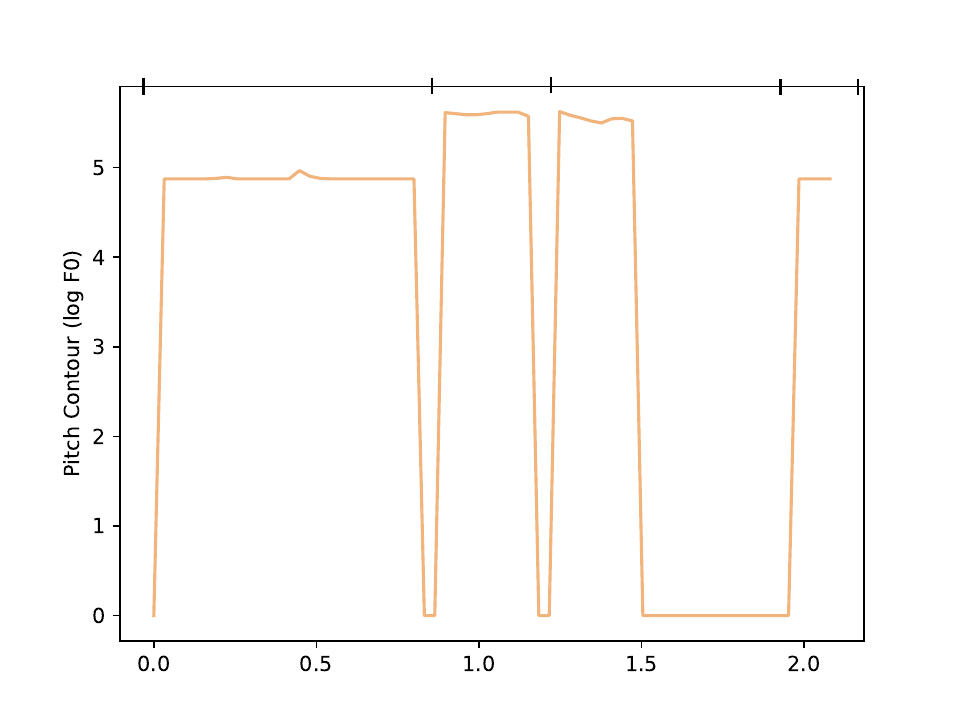}
    \label{fig:mel-c}
  }
  \subfigure[SCA result]{
    \includegraphics[width=0.22\textwidth]{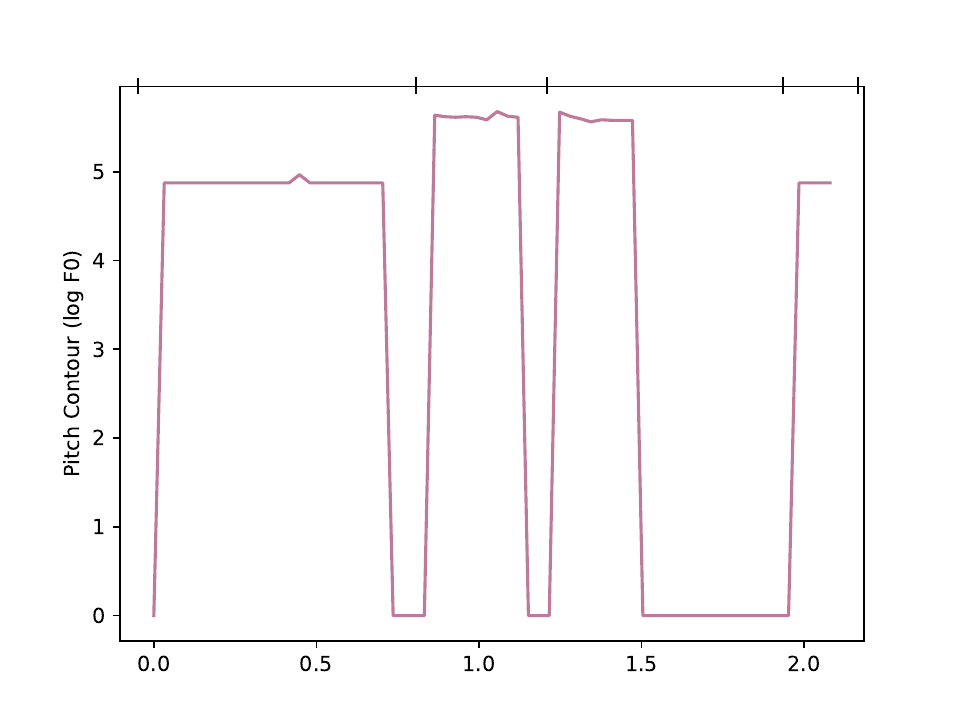}
    \label{fig:mel-d}
  }
  \caption{Pitch contours of the source audio, the style audio, the stylized audio, and the audio after SCA.}
  \label{fig:mel}
\end{figure}

\subsection{Acoustic Features Analyses}\label{sec:acoustic}
We conduct the evaluation of the two methods based on acoustic features, which extract audio waveforms for STA and pitch contours for SCA respectively.

Figure~\ref{fig:wav} displays the waveform of the original audio, the audio after style transfer, and after STA. STA indeed makes the perturbation more consistent with the original waveform. In contrast to traditional attack methods, applying style transfer results in significant changes in the waveform, posing a greater threat to ASR models. Specifically, Figure~\ref{fig:wav-b} and Figure~\ref{fig:wav-c} show that it is equivalent to applying traditional adversarial attacks on stylized audio, which only adds some perturbations to the audio waveform without altering its basic shape. However, as shown in Figure~\ref{fig:wav-a} and Figure~\ref{fig:wav-c}, our complete attack, due to the inclusion of the style transfer step, namely the transition from Figure~\ref{fig:wav-a} to Figure~\ref{fig:wav-b}, leads to significant changes in the shape of the waveform.

On the other hand, the SCA method directly acts on the style code, without generating noticeable noise. Moreover, with fewer iterations, the style of the resulting audio does not undergo significant changes, as shown in Figure~\ref{fig:mel}. Pitch is an important component of intonation and is one of the common style types in speech. This style feature is characterized by the shape of the pitch contour in the figure.
Rhythm describes how fast a speaker utters. Each sentence is divided into segments corresponding to each continuous pronunciation, marked on the top horizontal edge of the figure. Therefore, the lengths of these segments reflect the rhythm information.
As shown in Figure~\ref{fig:mel-a}, Figure~\ref{fig:mel-b}, and ~\ref{fig:mel-c}, style transfer successfully transfers the style of the style audio to the source audio. Figure~\ref{fig:mel-c} and Figure~\ref{fig:mel-d} show that the pitch contours are basically the same and there is no significant difference in the lengths of the pronunciation segments, indicating that our attack effectively preserves the characteristics of the original style code.

\begin{figure*}[t]
  \centering
  \subfigure[Remove pitch]{
    \includegraphics[width=0.31\textwidth]{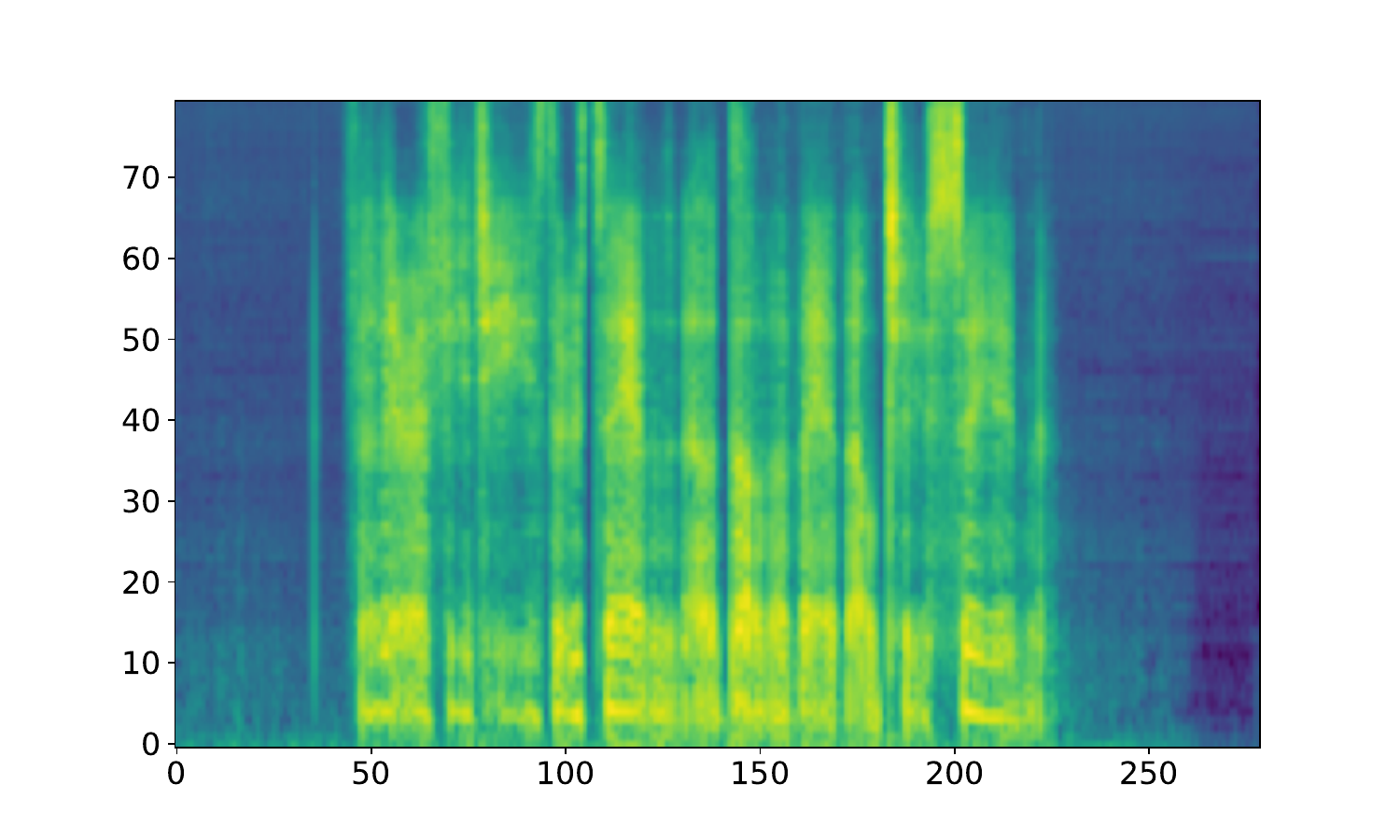}
    \label{fig:rm-a}
  }
  \subfigure[Remove rhythm]{
    \includegraphics[width=0.31\textwidth]{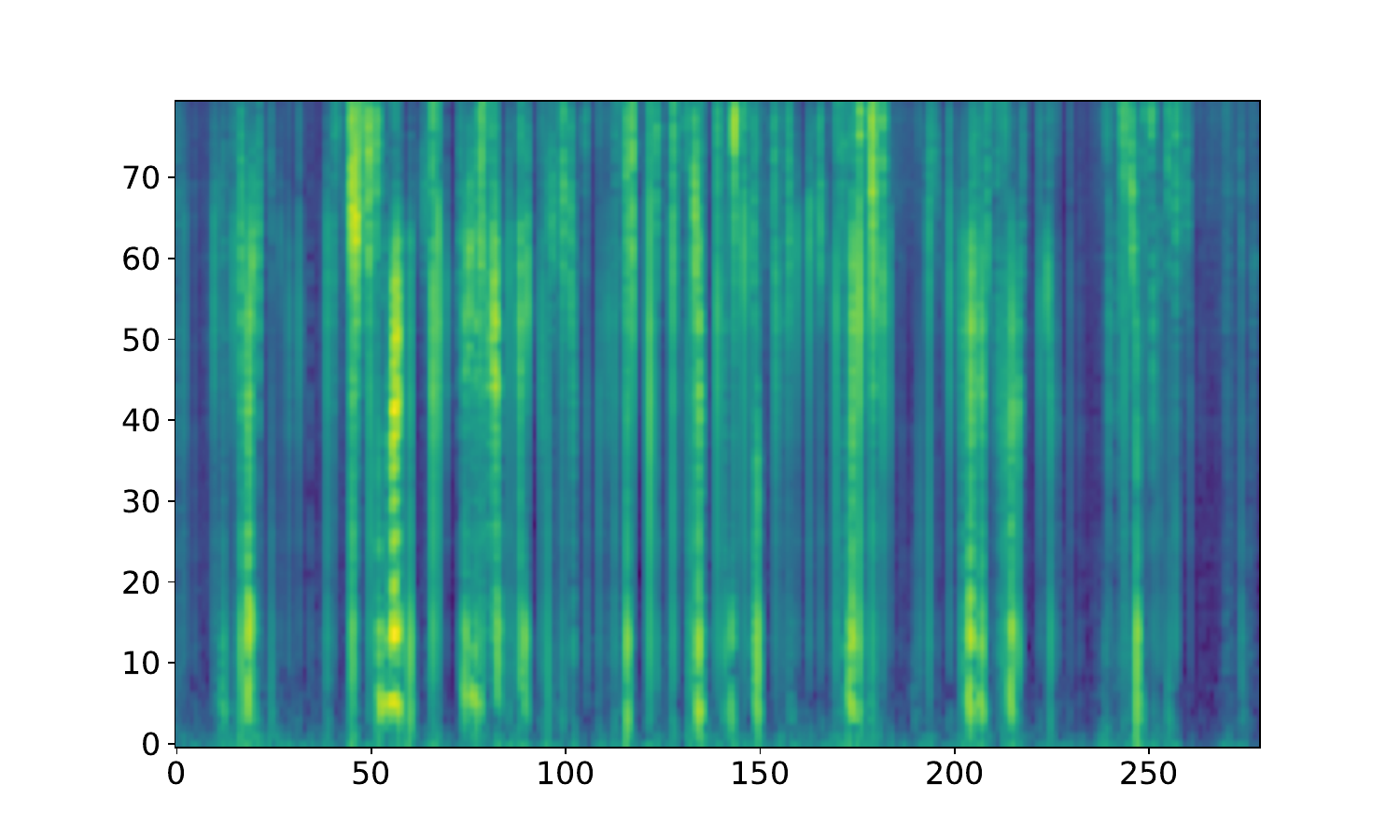}
    \label{fig:rm-b}
  }
  \subfigure[Remove both pitch and rhythm]{
    \includegraphics[width=0.31\textwidth]{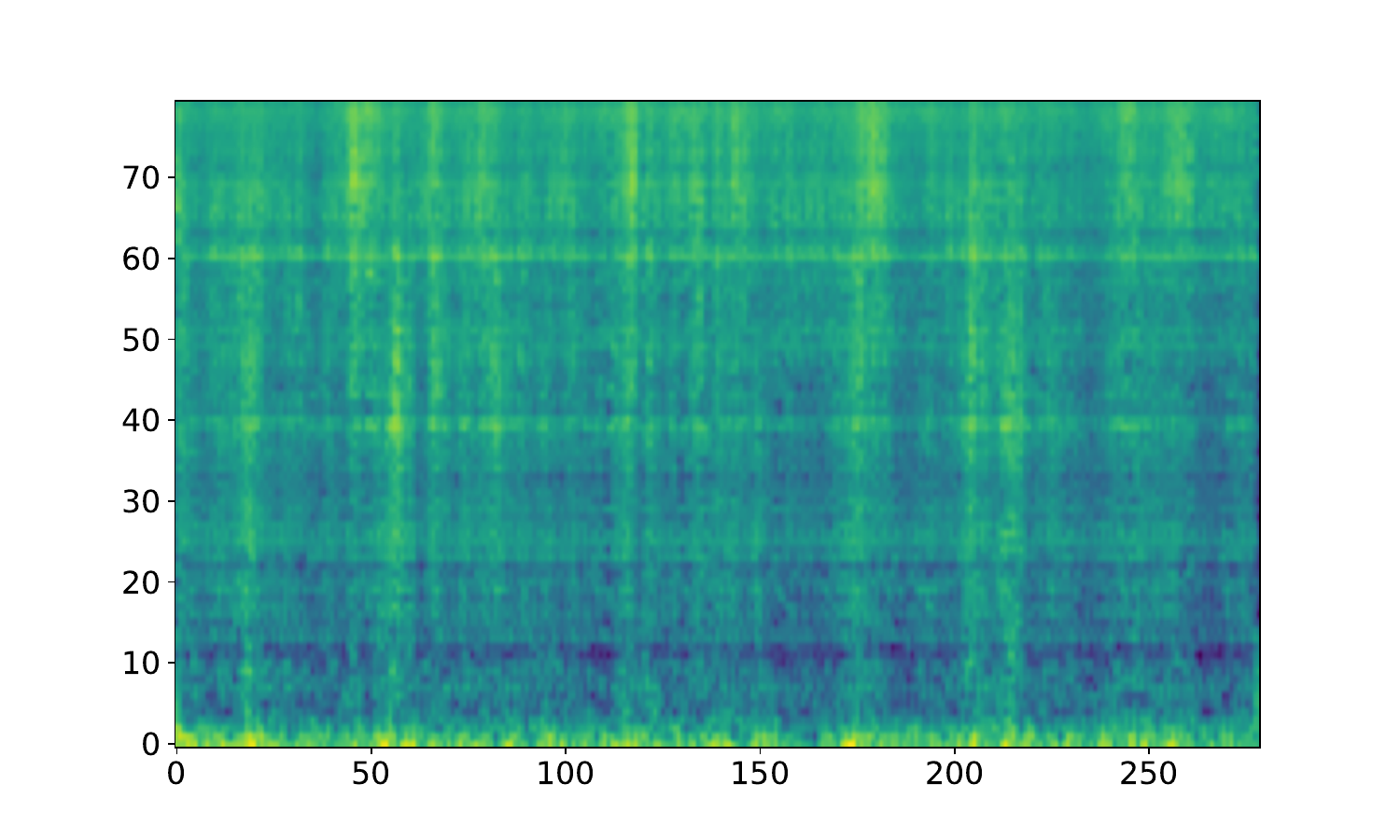}
    \label{fig:rm-c}
  }
  \caption{Spectrograms of the source audio after removing pitch, rhythm, both pitch and rhythm by initializing the corresponding style codes randomly.}
  \label{fig:remove}
\end{figure*}

Through extensive experiments, we do not find a suitable range of perturbations for the style codes. However, we find experimentally that SCA requires an average of 623 iterations for successful attacks, and failed examples tend to converge within fewer iterations. Within these iterations, the changes in the style codes are not sufficient to affect human perception. Our conclusion is that within a limited range, perturbations to the style codes do not significantly affect the style, and on this basis, successful attacks can be achieved. However, this does not mean that there are no cases where changes in the style codes lead to changes in style. As shown in Figure \ref{fig:remove}, by eliminating the style code, \ie initializing it randomly, we observe significant changes in style. Specifically, when the pitch is eliminated, there is a noticeable change in the tone of the resulting audio, but the audio remains within the perceptible range of human hearing. However, when the rhythm is eliminated, the resulting audio no longer contains precise semantic information perceivable by human hearing.

\subsection{Ablation Study}\label{sec:ablation}
To show the effectiveness of style transfer on pitch and rhythm, we first separately test the performance of STA and SCA by only transferring pitch (iteratively optimizing only the pitch code) or only transferring rhythm (iteratively optimizing only the rhythm code), referred to as ``STA-pit'', ``SCA-pit'', ``STA-rhy'', and ``SCA-rhy'', respectively. Then, we test the performance of applying the attack steps in STA without style transfer, referred to as ``Attack-only''. The specific results are shown in Table~\ref{tab:results}.

\noindent\textbf{Objective Evalution.} Comparing STA with the ``Attack-only'' me-thod, the average success rate of STA (74.7\%) is significantly higher than the success rate of the ``Attack-only'' method (61\%). Moreover, among the three STA style transfer schemes, even the lowest success rate, achieved in the scheme where only pitch is transferred (61\%) is higher than that of the ``Attack-only'' method. Additionally, the scheme involving both pitch and rhythm transfer achieves a success rate of 85\%, surpassing the success rates of schemes involving only pitch or only rhythm transfer. These results to some extent demonstrate the beneficial impact of style transfer on the attack.

\noindent\textbf{Subjective evalution.}
As shown in Figure~\ref{fig:MOS}, the ``Attack-only'' method yields similar audio quality to STA. However, in the six experiments involving style transfer, SCA consistently outperforms STA in terms of the MOS. SCA achieves an average MOS of 3.68, while STA achieves only 2.10 on average.

Furthermore, SCA's MOS is notably higher than that of style transfer without attack (3.68 compared to 3.50), indicating that SCA does not significantly degrade audio quality and naturalness. However, when comparing the MOS of benign audio (4.35) with that of style-transferred audio (3.50), the quality of the audio is influenced by the style transfer model itself. Therefore, achieving higher audio quality requires an improved style transfer model.

\section{Discussion}~\label{sec:discussion}

\noindent\textbf{Style Code.}
As shown in Table~\ref{tab:results}, the method of SCA iterating only the rhythm still maintains a relatively high success rate, but the success rate drastically decreases when iterating only the pitch. This suggests that perturbations in rhythm have a greater impact on the ASR model's understanding of semantic content in the audio. However, why perturbations in the pitch code do not significantly affect the semantic content of the audio remains an open question. This also leaves room for exploration in the interpretability of style codes, which we consider as a direction for our future work.

\noindent\textbf{Timbre Style Transfer.}
Timbre transfer may also be a direction worth exploring. Based on a model that generates stable speaker embeddings, iterating over speaker embeddings and constructing timbre using a well-trained decoder that supports zero-shot voice conversion may lead to ASR transcription errors.

\noindent\textbf{Potential Defense.}
Due to style transfer and previous attacks~\cite{qu2022synthesising,yu2023smack} that do not alter the content of the audio but modify other components, a potential defense strategy is ``de-stylization'', which involves converting all non-content parts of the audio into a specific style before entering the ASR model. This renders any style transfer attempts by attackers ineffective, ensuring accurate transcription when the audio is input into the ASR model.

\noindent\textbf{Audio Quality.}
Although we find that altering the style code does not significantly change the style of the generated audio, we still observe cases where the quality of the audio is affected in our experiments. Moreover, the quality of the adversarial examples we generate heavily relies on the quality of the audio produced by the style transfer model. How to improve the quality of the generated adversarial audio examples while ensuring that the transferred style remains unchanged is still an open challenge.

\section{Conclusion}
In this paper, we propose a novel approach to audio adversarial attacks based on style transfer, which includes two methods: style transfer attack and style code attack. The former utilizes style transfer to create a broad space of variations in the audio waveform, followed by adding subtle perturbations to mislead speech transcription. The latter iterates over style codes to achieve better speech quality and naturalness without significantly reducing the attack success rate. Our experiments demonstrate a significant threat to ASR models. Additionally, we observe a noticeable difference in the impact of iterating over pitch code or rhythm code on speech recognition during the experiments, inspiring further exploration into the specific meanings of style codes in the future. We hope that our work will inspire better defenses for ASR models to enhance their robustness and security.

\noindent \textbf{Ethical considerations.~}
The study underwent scrutiny by the Human Research Ethics Committee at the authors' institution, which concluded that it warranted exemption from further review involving human subjects. Adhering to ethical guidelines, we ensured the implementation of best practices in conducting surveys involving human participants. All individuals involved in the study were adults aged 18 and above, and they willingly consented to the utilization of their responses for academic research purposes.

\section{Acknowledgement}
The authors are grateful to the anonymous reviewers for their feedback that helped improve the paper. This research is supported in part by the National Natural Science Foundation of China (61801049).

{
    \small
    \bibliographystyle{unsrt}
    \bibliography{egbib}
}

\end{document}